\documentclass[12pt]{article}
\usepackage{amsmath,amsfonts,amssymb,graphicx,epsfig}
\usepackage{epsfig}
\date{\today}
\def\unit{\leavevmode\hbox{\small1\kern-3.6pt\normalsize1}}
\newcommand{\be}{\begin{equation}}
\newcommand{\ee}{\end{equation}}
\newcommand{\bea}{\begin{eqnarray}}
\newcommand{\eea}{\end{eqnarray}}

\def \ben{\begin{enumerate}}
\def \een{\end{enumerate}}
\def \bit{\begin{itemize}}
\def \eit{\end{itemize}}

\def\lsim{\raise0.3ex\hbox{$\;<$\kern-0.75em\raise-1.1ex\hbox{$\sim\;$}}}
\def\gsim{\raise0.3ex\hbox{$\;>$\kern-0.75em\raise-1.1ex\hbox{$\sim\;$}}}

\def \av#1{\left\langle #1\right\rangle}

\def \vckm{V_{\mathrm{CKM}}}

\def \GeV{{\mathrm{GeV}}}

\def \Im{{\mathrm{Im}}\,}
\def \Re{{\mathrm{Re}}\,}
\def \Tr{{\mathrm{Tr}}\,}

\def \diag{{\mathrm{diag}}}

\def \hc{\mathrm{H.c.}}

\def\21{$SU(2) \ot U(1)$}
\def\ot{\otimes}

\def\bold#1{\setbox0=\hbox{$#1$}
     \kern-.025em\copy0\kern-\wd0
     \kern.05em\copy0\kern-\wd0
     \kern-.025em\raise.0433em\box0 }
\def\arnps#1#2#3{{\it Annu. Rev. Nucl. Part. Sci.\/} {\bf#1} (#2) #3}
\def\epjc#1#2#3{{\it Eur. Phys. J.\/} {\bf C #1} (#2) #3}

\def\ibid#1#2#3{\emph{ibid.} {\bf #1} (#2) #3}

\def\nc#1#2#3{{\it Nuovo Cim.}~{\bf#1} (#2) #3}
\def\npb#1#2#3{{\it Nucl.~Phys.\/}~{\bf B #1} (#2) #3}

\def\plb#1#2#3{{\it Phys.~Lett.\/}~{\bf B #1} (#2) #3}

\def\prd#1#2#3{{\it Phys.~Rev.\/}~{\bf D#1} (#2) #3}
\def\prl#1#2#3{{\it Phys.~Rev.~Lett.\/}~{\bf #1} (#2) #3}

\def\zpc#1#2#3{{\it Z.~Phys.\/}~{\bf C #1} (#2) #3}
\def\hpph#1{{\tt hep-ph/#1}}

\def\hpex#1{{\tt hep-ex/#1}}

\def\jhep#1#2#3{{\it J.~High Energy Phys.}~{\bf #1} (#2) #3}
\begin{document}
\begin{flushright}
\small
FISIST/14-2001/CFIF\\
IPPP/01/58\\
DCPT/01/114\\
\end{flushright}
\renewcommand{\thefootnote}{\fnsymbol{footnote}}
\renewcommand{\baselinestretch}{1.2} \large\normalsize
\vspace{.3cm}
\begin{center}
{\bf{\Large 
Supersymmetry and a rationale for small CP \\ violating phases}}\\
\vspace* {1cm}
{\large 
G.C. Branco$^\mathrm{a}$,
M.E. G{\'o}mez$^\mathrm{a}$, 
S. Khalil$^\mathrm{b,c}$ and 
A.M. Teixeira$^\mathrm{a}$}\\
\vspace* {5mm}
$^\mathrm{a}$ {\it Centro de F\'\i sica das Interac{\c c}{\~o}es 
Fundamentais (CFIF),
Departamento de F{\'\i}sica,  Instituto Superior T{\'e}cnico, Av.
Rovisco Pais,  1049-001 Lisboa, Portugal}\\
\vspace* {2mm}
$^\mathrm{b}$ {\it IPPP, Physics Department, Durham University, DH1 3LE,
Durham, U.~K}\\
\vspace* {2mm}
$^\mathrm{c}$ {\it Ain Shams University, Faculty of Science, 
Cairo, 11566, Egypt}

\vspace*{10mm}
{\bf \large Abstract}
\end{center}

We analyze the CP problem in the context of a supersymmetric extension
of the standard model with universal strength of Yukawa couplings.
A salient feature of these models is that the CP phases are constrained to
be very small by the hierarchy of the quark masses and the pattern of CKM 
mixing angles. This leads to a small amount of
CP violation from the usual KM mechanism and a significant contribution
from supersymmetry is required. Due to the large generation mixing in 
some of the supersymmetric interactions, 
the electric dipole moments impose severe constraints on
the parameter space, forcing the trilinear couplings to be factorizable in
matrix form. We find that the $LL$ mass insertions give the dominant
gluino contribution to saturate $\varepsilon_K$. The chargino
contributions to $\varepsilon^\prime/\varepsilon$ are significant and
can accommodate the experimental results. In this framework,
the standard model gives a negligible contribution to the CP asymmetry in
B--meson decay, $a_{J/\psi K_S}$. However, due to supersymmetric 
contributions to $B_d-\bar{B}_d$ mixing, the recent large 
value of $a_{J/\psi K_S}$ can be accommodated.

\newpage
%
%
\section{\bf \large  Introduction}\label{intro}
The understanding of the origin of fermion families and the
observed pattern of fermion masses and mixings, together with the
origin of CP violation, are among the major outstanding problems
in particle physics. In the standard model (SM), since the flavour
structure of the Yukawa couplings is not defined by the gauge
symmetry, fermion masses and mixings are arbitrary parameters, to
be fixed by experiment.
A salient feature of the pattern
of quark masses and mixings is the fact that the spectrum is
strongly hierarchical and the mixings are small.

Concerning CP violation, it arises in the SM from complex
Yukawa couplings which lead to a physical $\delta_{CKM}$ phase in
the Cabibbo-Kobayashi-Maskawa (CKM) matrix. The strength of CP
violation in the SM is given by $\Im |Q_{CP}|$, with $Q_{CP}$ denoting any
rephasing invariant quartet of the CKM matrix. If one considers the
quartet $Q_{CP}=(V_{us} V_{cb} V_{ub}^* V_{cs}^*)$, the experimental
value of the moduli entering in this quartet constrains $|Q_{CP}|$ to be
of order $10^{-5}$, while the observed strength of CP violation in the
kaon sector requires $\phi \equiv \arg(V_{us} V_{cb} V_{ub}^* V_{cs}^*)$
to be of order one. Although in the SM no explanation is provided for
the experimentally observed values of $|V_{ij}|$, once these values are
incorporated, the SM accommodates in a natural way the observed
strength of CP violation in the kaon sector, measured by the parameter
$\varepsilon_K$. On the other hand, it has been established~\cite{bau}
that the strength of the CP violation in the SM is not sufficient to
generate the observed size of the baryon asymmetry of the
Universe. This provides the strongest motivation to consider new
sources of CP violation, beyond the one present in the SM.

In supersymmetric
(SUSY) extensions of the SM there are additional sources of CP
violation, due to the presence of new CP violating phases.
However, these new phases are severely constrained by experimental
bounds on electric dipole moments (EDMs)~\cite{phase:edm,savoy}. Setting these
new phases to zero is not natural in the sense of 't Hooft~\cite{hooft} since
the Lagrangian does not acquire any new symmetry in the limit
where these new phases vanish. This is the so-called SUSY CP-problem.
Since the question of CP-violation is closely related to the
general flavour problem, one may wonder whether it is possible,
within a supersymmetric extension of the SM, to establish a connection
between the need for small CP violating phases and the observed
pattern of quark masses and mixings.

Such a connection is possible if one
imposes universality of strength for Yukawa couplings (USY) on 
a SUSY extension of the SM. The idea of USY consists of assuming that
all Yukawa couplings have the same modulus, the flavour dependence
being all contained in the phases~\cite{branco:1990-1997}. 
This leads to what is known as pure phase mass matrices~\cite{hung:1998}.
This class of couplings can be motivated by local horizontal
symmetries (see Appendix),  and
it has also been recently shown that such textures may 
naturally arise within the framework of
a theory with two ''large'' extra dimensions~\cite{hung:2001}.

The fact that the quark masses are strongly hierarchical, 
together with the smallness of $|V_{cb}|$, $|V_{ub}|$ implies that
these phases have to be small, at most of order 
$m_2/m_3$~\cite{branco:1990-1997,branco:1995}.
One is then led to a scenario where the strength of CP violation
through the Kobayashi-Maskawa (KM) mechanism is naturally small, in
fact too small to account for the observed CP violation (CPV) 
in the kaon sector.
This provides motivation for embedding the USY ansatz in a larger
framework, where new sources of CP violation naturally arise.

In this paper, we impose the USY ansatz in the framework of SUSY. This
extension is specially interesting since 
new SUSY contributions to CP violation in the kaon sector may play an
important r\^ole in providing sufficient CP violation to saturate the
experimental value of $\varepsilon_K$.

We study the question of electric dipole moments (EDMs) in 
supersymmetric USY models.
Although the CP violating phases of the Yukawa couplings 
are small, the USY ansatz implies a large generation mixing in some of
the SUSY interactions, which in turn leads 
to potentially large contributions to the EDMs. The recent bound
from the mercury atom EDM~\cite{edm:hg} sets
important restrictions on the class of models we are considering.
We will show that due to these constraints one can only saturate the
observed value of $\varepsilon_K$ with non-universality of the 
soft trilinear couplings.

Finally, we analyze CP violation in the B-sector. 
Both the BaBar~\cite{2beta:babar} and Belle~\cite{2beta:belle} 
collaborations have provided clear evidence for CP
violation in the B--system, although at present the experimental errors
are relatively large. 
The large CP asymmetries observed by BaBar and Belle are in rough
agreement with the SM predictions. However, there is still quite a
large room for new physics and models with small CP violating 
phases like USY, as well as some of the models with CP as an approximate 
symmetry~\cite{approxCP} might remain phenomenologically viable, 
provided there are new physics contributions to $\varepsilon_K$,
$\varepsilon'/\varepsilon$, and $B_d-\bar{B}_d$
mixing~\cite{branco:1999}. 

This paper is organized as follows. In Section 2 we present the main
features of the USY ansatz for the Yukawa couplings. We show that 
USY is preserved under the renormalization group (RG) evolution, and
give a specific example of a choice of USY phases that leads to a
correct quark mass spectrum and $|V^{\mathrm{CKM}}_{ij}|$.
In Section 3 we implement the USY hypothesis within a
supersymmetric extension of the standard model and investigate
how the EDM constrains the SUSY parameter space. The remaining of the
section is devoted to the study of the kaon system CPV
observables. In Section 4 we analyze 
the implications of small CPV phases for CP
asymmetries in $B_d$ meson decays.
Our conclusions are presented in Section 5.


\section{\bf \large Universal strength of Yukawa couplings}\label{usy}

In the absence of a fundamental theory of flavour, one is tempted
to consider some special pattern for the Yukawa couplings, in the
hope that they could naturally arise from a flavour symmetry,
imposed at a higher energy scale. 
As mentioned in the introduction, one of the simplest suggestions
for the structure of the Yukawa couplings consists of assuming that
they have a universal strength $g_y$, so that the mass matrices 
can be written as
$m_{ij}=g_y v/\sqrt{2} \exp(i \phi_{ij})$, where $i,j$ denote family
indices. In the limit of small phases, the USY matrices can be viewed
as a perturbation of the democratic type matrices~\cite{democratic}.

Within the USY framework, the quark Yukawa couplings can be
parametrised as follows:
\begin{equation}\label{usy:usy:yuyd}
 U_{ij} = \frac{\lambda_u}{3} \exp \left[i \Phi^u_{ij}\right] \quad
\mathrm{and} \quad D_{ij} = \frac{\lambda_d}{3} \exp \left[i
\Phi^d_{ij}\right] \;,
\end{equation}
where $\lambda_{u,d}$ are overall real constants, 
and $\Phi^{u,d}$ are pure phase matrices. By performing appropriate
weak-basis (WB) transformations the matrices $U$ and $D$ can, without
loss of generality, be put in the form
\begin{equation}\label{usy:usy:ansatz}
U = \frac{\lambda_u}{3} \left(
\begin{array}{ccc}
e^{i {p_u}} & e^{i {r_u}} & 1 \\
e^{i {q_u}} & 1 & e^{i {t_u}} \\
1 & 1 & 1
\end{array} \right) \;;
\quad \quad
D = \frac{\lambda_d}{3} K^\dagger \left(
\begin{array}{ccc}
e^{i {p_d}} & e^{i {r_d}} & 1 \\
e^{i {q_d}} & 1 & e^{i {t_d}} \\
1 & 1 & 1
\end{array} \right) K \;,
\end{equation}
where $K=\diag(1, e^{i \kappa_{1}},e^{i \kappa_{2}})$.

The phases $\kappa_i$ do not affect the spectrum of quark masses, but
influence $\vckm$. The phases ${(p,q,r,t)}$ affect both the
spectrum and $\vckm$. As previously remarked, 
the hierarchical character of the quark
masses constrains each of these latter phases to be small. In order to
see how these constraints arise, it is useful to define the Hermitian
matrices $H_{u,d} = 3/\lambda_{u,d}^2\; Q Q^\dagger$, 
where $Q=U,D$. It can then
be readily verified that the second invariant of $H$ is given
by~\cite{branco:1990-1997}
\begin{equation}\label{usy:usy:chiH}
\chi(H)= 
\frac{4}{9} \sum_m \sin^2 \sigma_m = 
9\;\frac{m_1^2 m_2^2 +m_1^2 m_3^2 +m_2^2 m_3^2}{{(m_1^2+m_2^2+m_3^2)}^2}\;,
\end{equation}
where, in each sector,
$\sigma_m$ are linear combinations of the phases ${(p,q,r,t)}$.
Since $\chi(H)$ is a sum of positive definite quantities, the modulus
of each of the phases has to be small, at most of order
$m_2/m_3$.
As mentioned above, the phases $\kappa_i$ do not affect the quark
spectrum. However, it has been shown~\cite{branco:1995} that the fact that 
${(|V_{cb}|^2+|V_{ub}|^2)}$ is small, constrains each one of the
phases $\kappa_1, \kappa_2$ to be at most of order $|V_{cb}|$. 
As we will see, these $\kappa$ phases will prove to be essential to
control the size of the $\vckm$ elements, in particular that
of $V_{td}$.

In our search for a choice of phases leading to the correct masses and
mixings, we expand both $\chi(H)$ and the determinant of $H$ in the
limit of small phases, and per each of 
the quark sectors, we write two of the phases as
functions of the masses and of the  
remaining two phases. At this point, we have still some
freedom, which we use to fit the experimental bounds on the $\vckm$ matrix.
Therefore, for fixed values of the quark 
masses\footnote{At $\mu=m_{top}$, 
we use the following values for the running quark
masses: $m_u=0.0018\; \GeV$, $m_d=0.0048\; \GeV$, $m_s=0.075 \;\GeV$, 
$m_c=0.6 \;\GeV$, $m_b=2.9 \;\GeV$, $m_t=170 \;\GeV$, consistent 
with the values given in Ref.~\cite{koide:1998}.} 
and using as
input the  phases $p_{u,d}$, $r_{u,d}$ and $\kappa_i$, we find regions of the
phase-parameter space where one correctly reproduces the masses and
mixings of the quarks. 

As referred in the introduction, USY can be generated at some
high energy scale through the breaking of a flavour symmetry. 
In Ref.~\cite{branco:1990-1997}, 
it was shown that USY can account for the quark mass
spectrum at low energy scales and thus it is important to investigate whether
this structure for the Yukawa couplings is preserved by the
renormalization group equations (RGE)~\cite{rge:bbo}.

At the weak scale, the phases required to 
explain the fermion masses and mixings are
small, hence we can expand the
Yukawa couplings of Eq.~(\ref{usy:usy:yuyd}) as
\begin{equation}\label{usy:rge:y-re-i}
U_{ij} = \frac{\lambda_u}{3} \left[E_{ij}+ i \Phi^u_{ij}\right] \quad 
\mathrm{and} \quad 
D_{ij} = \frac{\lambda_d}{3} \left[E_{ij} + i \Phi^d_{ij}\right] \;.
\end{equation}
In the above equation, we neglected terms
$\mathcal{O}(\Phi^2_{q})$, and $E$ is the democratic matrix, defined as 
$E_{ij}=1, \; \forall i,j$ .
The RGE for the real part of Eq.~(\ref{usy:rge:y-re-i}) can be easily
identified with the evolution 
of the top and bottom Yukawa couplings:
\begin{align}\label{usy:rge:rgeup-real}
& \frac{d \; \lambda_u}{dt} = \frac{1}{16 \pi^2} 
\left( -G_u + 6 \lambda_u^2 +\lambda_d^2 \right) \lambda_u 
\;,\nonumber \\
&\frac{d \; \lambda_d}{dt} = \frac{1}{16 \pi^2} 
\left( -G_d + 6 \lambda_d^2 + \lambda_u^2 +
\lambda_l^2 \right) \lambda_d \;,
\end{align}
where $G_k(t) = \sum_{i=1}^{3} c^i_k \; g^2_i(t)$, with 
$g_i(t)$ the running gauge couplings for 
$SU(3) \times SU(2) \times U(1)$ gauge
group and $c_k$ the MSSM RGE coefficients.
The reason why, in this limit, the running of the overall Yukawa
coefficient decouples from the running of the phases
becomes clear if, by means of a weak basis
transformation, one moves
from the USY basis to the ``heavy" basis
$(Y \to Y^\prime=F^\dagger Y F)$, where 
\begin{equation}\label{usy:rge:heavy}
Y_{u,d}^\prime=\frac{\lambda_{u,d}}{3} 
\left[\diag(0,0,3) + 
i \varphi^{u,d}_{ij}\right] \;,\;\mathrm{with}
\end{equation}
\begin{equation}\label{usy:rge:heavy-rotF}
\varphi^k(t) = F^\dagger \Phi^k  F  \quad \mathrm{and} \quad
F= \left[
\begin{array}{ccc}
  1/\sqrt{2} & -1/\sqrt{6} & 1/\sqrt{3} \\
  0          &  2/\sqrt{6} & 1/\sqrt{3} \\
 -1/\sqrt{2} & -1/\sqrt{6} & 1/\sqrt{3} \\
\end{array} 
\right] \; .
\end{equation}
The lighter generation Yukawa couplings are thus 
small perturbations of the leading third generation, originated by
the small USY phases.

The running of the imaginary parts is 
\begin{align}\label{usy:rge:rgeup-phase}
\frac{d \; \Phi^u}{dt} &= \frac{1}{16 \pi^2} \left[ \lambda_u^2 (-3 I + E) 
+ \frac{\lambda_d^2}{3} (-3 I+ E) \right] \Phi^u 
+ \frac{1}{16 \pi^2} \left( \frac{\lambda_u^2}{3} \Delta^u +
\frac{\lambda_d^2}{9}  \Delta^d \right) E \; ,\nonumber \\
\frac{d \; \Phi^d}{dt} &= \frac{1}{16 \pi^2} \left[ 
\lambda_d^2  (-3 I + E) 
+ \frac{\lambda_u^2}{3}  (-3 I+ E) \right] \Phi^d + 
 \frac{1}{16 \pi^2} \left(\frac{\lambda_d^2}{3}\;  
\Delta^d  E +
\frac{\lambda_u^2}{9}  \Delta^u  E \right) \;,
\end{align}
where $I$ denotes the $3 \times 3$ identity matrix
and $\Delta^k = \Phi^k E - E ({\Phi^{k}})^T$.
The scale dependence of the phases can be explicitly analyzed
if we write Eqs.~(\ref{usy:rge:rgeup-phase}) in the ``heavy'' basis:
\begin{equation}\label{usy:rge:psiut}
\frac{d \; \varphi^u}{dt} = -(3 k_u^2 + k_d^2) 
\left[
\begin{array}{ccc}
  \varphi^u_{11} & \varphi^u_{12} & 0 \\
  \varphi^u_{21} & \varphi^u_{22} & 0 \\
       0         &        0       & 0 \\
\end{array}
\right] - k_d^2 \left[
\begin{array}{ccc}
  0 & 0 & \varphi^u_{13}-\varphi^d_{13} \\
  0 & 0 & \varphi^u_{23}-\varphi^d_{23} \\
  0 & 0 &      0 \\
\end{array}  
\right] \;,
\end{equation}
\begin{equation}\label{usy:rge:psidt}
\frac{d \; \varphi^d}{dt} = -(3 k_d^2 +  k_u^2) 
\left[
\begin{array}{ccc}
  \varphi^d_{11} & \varphi^d_{12} & 0 \\
  \varphi^d_{21} & \varphi^d_{22} & 0 \\
       0         &        0       & 0 \\
\end{array}
\right] - k_u^2 \left[
\begin{array}{ccc}
  0 & 0 & \varphi^d_{13}-\varphi^u_{13} \\
  0 & 0 & \varphi^d_{23}-\varphi^u_{23} \\
  0 & 0 &      0 \\
\end{array}  
\right] \;,
\end{equation}
where we defined $k^2_j(t) = \lambda_j^2(t)/16 \pi^2$.

We have performed the analysis of the scale dependence of the phases 
in the framework of the SM and in the MSSM. The differential
equations were solved numerically and analytically, the latter in the limit of
small phases. In both cases we have found that 
if the USY phases are small at some initial scale (which in a
phenomenologically consistent model is justified by 
the smallness of the phases required to 
fit the quark masses and the $\vckm$ at the weak scale), 
then the approximation of Eq.~(\ref{usy:rge:y-re-i})
remains valid to energies up to the GUT scale.

The explicit dependence of the phases on the scale can be easily  
displayed for the low $\tan \beta$ regime, where one can neglect terms in 
$\lambda_d^2$ and $\lambda_l^2$ relatively to the leading 
$\lambda_u^2$ contributions. In this limit, the evolution of the up
phases is given by
\begin{equation}\label{usy:rge:ltb:phiu}
\Phi^u(t) =a(t) \Phi^u(0) +(1-a(t))\chi^u \; ,
\end{equation}
where $a(t)= \exp \left(-3 \int^t_0 k_u^2(t') dt'\right)$ contains the scale 
dependence and 
\begin{equation}\label{usy:rge:ltb:chiu}
\chi^u = 
F \left[
\begin{array}{ccc}
 0 & 0 & \varphi^u_{13}(0) \\
 0 & 0 & \varphi^u_{23}(0) \\
  \varphi^u_{31}(0) & \varphi^u_{32}(0) & \varphi^u_{33}(0) \\
\end{array}
\right] F^\dagger  \;,
\end{equation}
is a function of the initial phases; 
the structure of $\chi^u$ allows these to be combined so that the 
up quark Yukawa coupling for low $\tan\beta$ values takes the form:
\begin{equation}\label{usy:rge:ltb:yuk-u-K}
U(t) = \frac{\lambda_u(t)}{3} \; {K_L^u}^\dagger \; 
e^{  i  a(t) \Phi^u(0) } \; K_R^u \;,
\end{equation}
where $K_L^u$ and $K_R^u$ are diagonal phase matrices, whose entries
depend on the initial phases contained in the matrix $\chi^u$ and on 
the integral $a(t)$.

The analysis for the down sector is entirely analogous. In the low 
$\tan \beta$ regime, the solution for the running of the down phases
takes the form: 
\begin{equation}\label{usy:rge:ltb:phid}
\Phi^d(t) =b(t) \Phi^d(0) +(1-b(t))\chi^d +
d(t) \eta^u \;,
\end{equation}
where $b(t)= \exp \left(- \int^t_0 k_u^2(t') dt'\right)$ and 
$d(t)=  -b(t)  \int^t_0 \frac{k_u^2(t')}{b(t')} dt'$. The matrices 
$\chi^d$ and $\eta^u$ are also functions of the initial phases.
\begin{equation}\label{usy:rge:ltb:chid-eta}
\chi^d = 
F \left[ 
\begin{array}{ccc}
 0 & 0 & 0 \\
 0 & 0 & 0 \\
 \varphi^d_{31}(0) & \varphi^d_{32}(0) & \varphi^d_{33}(0) \\
\end{array}
\right] F^\dagger  \;,\quad
\eta^u =  
F  \left[ 
\begin{array}{ccc}
0 & 0 & \varphi^u_{13}(0) \\
0 & 0 & \varphi^u_{23}(0) \\
0 & 0 & 0 \\
\end{array}
\right] F^\dagger  \;.
\end{equation}
Just like in the previous case, these phase matrices can be reorganised  
so that the down Yukawa coupling evolution reads:
\begin{equation}\label{usy:rge:ltb:yuk-d-KQ}
D(t) = \frac{\lambda_d(t)}{3} {K_L^d}^\dagger \;Q \; 
e^{  i  b(t) \Phi^d(0) } \; K_R^d \;,
\end{equation}
where $K_L^d, \ K_R^d $ and $Q$ are diagonal phase matrices; 
the first two depend on the function $b(t)$ and on the phases
appearing on $\chi^d$, while the third depends on $d(t)$ and 
$\eta^u$.

By means of yet another weak basis transformation, one can still
rewrite the Yukawa couplings in 
Eqs.~(\ref{usy:rge:ltb:yuk-u-K},\ref{usy:rge:ltb:yuk-d-KQ}) as
\begin{equation}\label{usy:rge:ltb:yuk-Krot}
U(t) \; \rightarrow \; \frac{\lambda_u(t)}{3} \; 
e^{  i a(t) \Phi^u(0) }\;;  \quad \quad
D(t) \; \rightarrow \; \frac{\lambda_d(t)}{3} \; P^\dagger \; 
e^{  i b(t) \Phi^d(0) } \; P   
\quad \quad (P=Q^\dagger K^d_L {K^u_L}^\dagger)
\; .
\end{equation}
This means that it is always possible to move to a basis where 
the Yukawa couplings are easily related to the original ones.
Obviously, the $P$ matrices have no effect on the quark spectrum,
only on $\vckm$. Therefore 
RG evolution preserves the USY texture that is responsible for the
non-degenerate quark spectrum.
For the case of large $\tan \beta$, the argument can be equally
verified.

At this point, one should make a 
final remark regarding the possibility of having Hermitian 
and USY Yukawa couplings, that is having the phase matrices
verify the condition $\Phi^q_{ij} = -\Phi^q_{ji}$, ($q=u,d$). 
As it was first pointed out in \cite{branco:1994}, 
Hermitian USY leads to a sum rule for the
quark masses, namely $m_1 \; m_2 +m_2\; m_3 +m_1\; m_3 = 0$,
which is clearly violated by experiment. 
The question that naturally follows is whether Hermitian 
USY textures imposed at some higher energy (GUT scale, 
for example) can produce a phenomenologically viable
texture through RG effects. As it can be seen from
Eqs.~(\ref{usy:rge:ltb:yuk-u-K},\ref{usy:rge:ltb:yuk-d-KQ},
\ref{usy:rge:ltb:yuk-Krot}),
if one starts with Hermitian structures, that is $\Phi^q_{ij}(0) =
-\Phi^q_{ji}(0)$, we find that at any other scale one can always
rotate to a basis where the phases that govern the spectrum 
scale with the flavour-independent coefficients $a(t)$ and $b(t)$, so 
that the impossibility of having Hermitian USY textures holds at any scale.

To illustrate some of the features of this ansatz, let us
consider, as an example, the following choice of USY phases:
\begin{eqnarray}\label{usy:ex:2}
U &=& \frac{\lambda_u}{3} \left(
\begin{array}{ccc}
e^{i \; 0.000517} & e^{-i \; 0.0000074} & 1 \\
e^{-i \; 0.01438} & 1 & e^{i\;  0.00158} \\
1 & 1 & 1
\end{array} \right) ,\; \nonumber\\
D &=& \frac{\lambda_d}{3} K^\dagger \left(
\begin{array}{ccc}
e^{i \; 0.00229} & e^{i \; 0.02833} & 1 \\
e^{-i \; 0.09278} & 1 & e^{-i \; 0.1453} \\
1 & 1 & 1
\end{array} \right) K \;,\nonumber \\
K &=& \diag (1\;,\;1\;,\;e^{i \; 0.0138}) \;.
\end{eqnarray}

The overall factor has been defined as 
$ \lambda_u/3 = m_{t}/v \sin \beta$ and 
$ \lambda_d/3 = m_{b}/v \cos \beta$.

The Yukawa couplings can be diagonalized by means of the
transformations
\begin{equation}\label{usy:ex:diag}
S_R^{d(u)}~ Y^{d(u)^T}~ S_L^{d(u)^{\dag}} = Y^{d(u)}_{\mathrm{diag}}\;,
\end{equation}
where $S_R$ and $S_L$ are unitary matrices. In the present USY model
we find 
\begin{equation}\label{usy:ex:sul}
S_L^u =  \left(
\begin{array}{ccc} 
0.695 & 0.0236-1.66 \times 10^{-4} \;i & -0.719+1.83 \times 10^{-4} \;i \\
-0.429 & 0.816-3.72 \times 10^{-3} \;i & -0.388+1.65 \times 10^{-4} \;i \\
0.577 i\; & 2.56 \times 10^{-3} + 0.577 \;i & 9.79 \times 10^{-5}+0.577 \;i
\end{array}\right)\;,
\end{equation}
\begin{equation}\label{usy:ex:sdl}
S_L^d =  \left(
\begin{array}{ccc} 
0.772 & -0.157 +0.0122\;i & -0.615-4.99 \times 10^{-4} \;i \\
-0.264 & 0.798-0.0759 \;i & -0.537+3.05 \times 10^{-3} \;i \\
0.578 \;i & 0.511 + 0.575 \;i & -2.07 \times 10^{-3}+0.577 \;i
\end{array}\right)\;,
\end{equation}
and the associated $\vckm$ 
\begin{equation}\label{usy:ex:vckm2}
\vckm = \left(
\begin{array}{ccc} 
0.975 -6.88 \times 10^{-4} \;i & 0.221+3.96 \times 10^{-3} \;i
& 0.0027 - 4 \times 10^{-5} \;i \\
-0.220 -9.71 \times 10^{-3} \;i & 0.973+0.061 \;i & 
0.041+1.92 \times 10^{-3} \;i \\
0.0063 + 1.10 \times 10^{-4} \;i & -0.041 -1.75 \times 10^{-3} \;i &
0.999+0.027 \;i
\end{array}\right)\;.
\end{equation}
Within the context of the SM, the matrices $S_{L,R}^{u,d}$ do not
have, by themselves, any physical meaning, only the combination 
$S_L^u . {S_L^{d^{\dagger}}} \equiv \vckm$ is physically meaningful. 
However the
matrices $S_{L,R}^{u,d}$ do play a significant r\^ole in some
extensions of the SM, as, for example the MSSM with non--universal
soft-breaking terms. We shall discuss this question in the following
sections.

The values of the moduli of the above presented 
$\vckm$ are in agreement with
the experimental bounds. The central value of $|V_{td}|$ 
($|V_{td}| \sim 0.0063$) is
somewhat smaller than the central value of $|V_{td}|$ extracted from 
experiment~\cite{PDG:2000}, within the context of the SM, namely 
$|V_{td}|=0.0073$. However, it should be noted that in the framework of
the SM, the value of $|V_{td}|$ is derived from the experimental value
of $B_d-\bar{B}_d$ mixing, and not from tree-level decays, as it is
the case for the moduli of the first two rows of $\vckm$. Since in the
SM, $B_d-\bar{B}_d$ mixing only receives contributions at one loop
level, there may be non-negligible contributions to  $B_d-\bar{B}_d$
mixing from physics beyond the SM.

Concerning CP violation, it
can be readily verified that the above $\vckm$ leads to a strength of
CP breaking (measured by 
$J\equiv \Im (V_{ud} V_{cs} V^*_{us} V^*_{cd})$~\cite{jarlskog:1985}) 
that is too small to account for the observed value of 
$\varepsilon_K$. In fact, for the above example, one obtains 
$J \sim 10^{-7}$, which is to be compared to the required value 
of $J \sim 10^{-5}$. This is a generic feature of the USY ansatz.
Although one can obtain somewhat larger values of
$J$ (e.g. $J \sim 10^{-6}$) for a different choice of USY 
phases~\cite{branco:1990-1997}, the tendency in this class of
ans\"atze is to have values of $J$ which are too small to saturate the
experimental value of $\varepsilon_K$. This motivates the embedding of
USY in a larger framework where new sources of CP violation naturally
arise. In the following sections we address this question in the
context of supersymmetric extensions of the standard model.

%
\section{\bf \large Supersymmetric USY models and CP violation}\label{susy}
As advocated in the introduction, supersymmetric extensions of
the SM may provide a considerable enhancement to CP violation
observables, since in addition to containing new ``genuine''
CPV phases, they introduce new flavour structures.
Accordingly, SUSY emerges as the natural candidate to solve the problem
of CP violation inherent to the USY models discussed in the
previous section.
We recall that within the USY framework 
the amount of CP violation arising from the CKM mechanism is typically
quite small,  
hence one will need significant supersymmetric contributions
that both succeed in saturating the values of the CP violating observables  
and in having EDM contributions which do not violate the 
current experimental bounds.

We will consider the minimal supersymmetric standard model (MSSM),
where a minimal number of superfields is introduced and $R$ parity
is conserved\footnote{For models with broken $R$ parity where universal
  strength of Yukawa couplings is also assumed, see for example 
Ref.~\cite{liu}.}, with the following soft SUSY breaking terms 
\bea\label{susy:gen:vsb}
V_{SB} &=& m_{0\alpha}^2 \phi_{\alpha}^* \phi_{\alpha} + 
\epsilon_{ab} 
(A^u_{ij} Y^u_{ij} H_2^b \tilde{q}_{L_i}^a \tilde{u}^*_{R_j} +
A^d_{ij} Y^d_{ij} H_1^a \tilde{q}_{L_i}^b \tilde{d}^*_{R_j} +
A^l_{ij} Y^l_{ij} H_1^a \tilde{l}_{L_i}^b \tilde{e}^*_{R_j} \nonumber\\
&-& B\mu H_1^a H_2^b + \mathrm{H.c.}) 
- \frac{1}{2} 
(m_3\bar{\tilde{g}} \tilde{g} + 
m_2 \overline{\widetilde{W^a}} \widetilde{W}^a +
m_1 \bar{\tilde{B}} \tilde{B})\;,
\eea 
where $i,j$ are family indices, $a,b$ are $SU(2)$ indices, and 
$\epsilon_{ab}$ is the $2\times 2$ fully antisymmetric tensor, with
$\epsilon_{12}=1$. Moreover, $\phi_{\alpha}$ denotes all 
the scalar fields of the theory.
Although in general the parameters $\mu$,
$B$, $A^\alpha$ and $m_i$ can be complex, two of their
phases can be rotated away~\cite{hall:85}. 
In a minimal SUGRA scenario (mSUGRA), the soft SUSY breaking
parameters are universal at some very high energy scale
(which we take to be the GUT scale), and we can write
\begin{equation}\label{susy:gen:msugra}
m_{0\alpha}^2 = m_0^2 \;, \quad m_i=m_{1/2}\;, \quad
A^\alpha_{ij} = A_0 e^{i \phi_A}\;.
\end{equation}
In this case, there
are only two physical phases 
\begin{equation}\label{susy:gen:msugraphase}
\phi_{A} = \arg(A^* m_{1/2}) , ~~~ \phi_{\mu} =
\arg(\mu ~m_{1/2}).
\end{equation}

In order to have EDM values below the experimental bounds, and
without forcing the SUSY masses to be unnaturally heavy, 
the phases $\phi_{A}$ and $\phi_{\mu}$ must be at most of order 
$10^{-2}$~\cite{phase:edm}. Thus, in this class of SUSY models with minimal 
flavour violation, 
complex Yukawa couplings leading to a physical $\delta_{CKM}$ phase in
the CKM matrix are the main source of 
CP violation. It should be also stressed that even if 
one ignores the bounds from the EDMs, and allows 
$\phi_A$ and $\phi_\mu$ to be of order one,
the latter models do not generate any sizeable new contribution to 
$\varepsilon_K$ and $\varepsilon^\prime/\varepsilon$~\cite{demir}
apart from those present in the SM. In the mSUGRA scenario of 
Eq.(\ref{susy:gen:msugra}), the USY structure of 
the Yukawa couplings is inherited by the trilinear terms so this 
simple flavour structure is excluded, since, and as aforementioned, no
new contributions to CPV observables are generated. 
However, in general  supergravity 
scenarios \cite{Brignole:1997dp}, the Yukawa couplings can be  
independent of the hidden sector fields that break supersymmetry and hence 
one can have $A$-terms with a flavour structure that is 
not related to that of the 
Yukawa couplings. This has recently motivated a growing interest on 
supersymmetric models 
with non--universal soft breaking terms, and a considerable amount of work 
has been devoted to the analysis of 
the effects of the new flavour textures on the 
CPV observables~\cite{susycp,vives,khalil}.

Regarding the non--universality of the trilinear soft terms, 
it has been emphasized that a new flavour structure in the $A$ terms
can saturate the experimental bounds on $\varepsilon_K$ and
$\varepsilon'/\varepsilon$ even in the presence of a vanishing 
$\delta_{CKM}$~\cite{susycp,vives}. However, the phases associated with 
the $A$--term diagonal elements might induce large EDMs. 
The latter problem can be overcome if one takes the Yukawa couplings and 
the $A$--terms to be Hermitian, in order to ensure
that the diagonal elements of $A_{ij} Y_{ij}$ are real in any 
basis~\cite{khalil,Khalil:2002jq}. 
This problem is also less severe if the trilinear
terms can be factorized as  $A_{ij} Y_{ij} = A.Y$ or
$Y.A$, an interesting possibility that naturally
arises withing the context of general SUGRA models~\cite{vives}.
As we will discuss below, this factorization implies that the 
mass insertion $(\delta_{11}^d)_{LR}$ (which is the most relevant  
to the EDMs contributions) is 
suppressed by the ratio $m_d/m_{\tilde{q}}$, where $m_d$ is the down quark
mass and $m_{\tilde{q}}$ is an average squark mass.

The non-universality of the squark mass matrices, which can be
arbitrary in generic SUSY models, is not constrained by the
EDMs. Nevertheless, $\Delta m_K$ and $\varepsilon_K$ impose 
severe constraints on the squark mixing, which can still be avoided in
models with flavour symmetries or squark-quark alignment~\cite{Masiero:2001cc}.

After this brief discussion, we begin the analysis of the implications
of having USY Yukawa couplings within a supersymmetric scenario of CP
violation. 
As in most SUSY models, the trilinear terms will play a key
r\^ole, since they are directly related with the 
increasingly more stringent bounds coming from the EDMs.  
In what follows, we will study
the enhancement of the CP observables through the non-universalities of 
the soft breaking terms in supersymmetric USY models.  

%
\subsection{\bf \normalsize
Constraints from the electric dipole moments}\label{susy:edm}
In this section we investigate how the EDM bounds constrain the
parameter space for SUSY models with universal strength of Yukawa 
couplings. The current experimental bound on the EDM of the neutron is given by
~\cite{edm:neutron}
\begin{equation}\label{susy:edm:dnexp}
d_n < 6.3 \times 10^{-26}\; e\ \mathrm{cm} \quad (90\% ~\mathrm{C.L.}).
\end{equation}
This bound can be translated into constraints for the imaginary parts of the 
flavour conserving $LR$ mass insertions.
\begin{equation}\label{susy:edm:dndeltaLR}
(\delta_{11}^{d(u)})_{LR} = \frac{1}{\tilde{m}_q^2} \left[
\left(S_L^{d(u)}~ Y^{A^*}_{d(u)}~ S_R^{d(u)^{\dag}}\right)_{11}
v_{1(2)} - \mu Y_1^{d(u)}~ v_{2(1)} \right],
\end{equation}
where $\tilde{m}_q^2$ is the average squark mass and
$v_i=\av{H^0_i}/\sqrt{2}$. In our analysis, the $\mu$ term is assumed
to be real and its magnitude is computed from electroweak symmetry breaking.
Recall that $S_L^{d(u)}$ and $S_R^{d(u)}$ are the rotation matrices
that diagonalize the down (up) Yukawa couplings.
For the specific case of USY couplings discussed in Section~\ref{usy}, 
the structure of $S_L^{d(u)}$ was presented in
Eqs.~(\ref{usy:ex:sul},\ref{usy:ex:sdl}).
$S_R^{d(u)}$ manifest the same large intergenerational mixing and 
large phases. The effects of these phases are absent in the
SM and in supersymmetric models with universal soft-breaking terms
(i.e., flavour structures that are aligned with the Yukawa couplings).
Nevertheless, the effect of any minimal deviation from this 
scenario, particularly
regarding the non-degeneracy of the trilinear terms, is strongly enhanced 
due to the large mixing, providing significant contributions to CP
violation observables. On the other hand, the bounds arising from 
the experimental
measurements of the EDMs severely constrain any non-degeneracy of the
trilinear terms, even in the limit of vanishing SUSY phases.
Imposing that $d_n$ does not exceed the experimental limit 
requires $\Im (\delta^{d(u)}_{11})_{LR}$ to be less than $10^{-6}$.
However, a stronger constraint on these mass insertions comes from the
recently measured electric dipole moment of the mercury
atom~\cite{edm:hg} 
\begin{equation}\label{susy:edm:dhgexp}
d_{Hg} < 2.1 \times 10^{-28}\; e \ \mathrm{cm}\;.
\end{equation}
This constraint corresponds to having
$\mathrm{Im}(\delta^{d(u)}_{11})_{LR} \lsim 10^{-7}
-10^{-8}$. Since the mercury EDM also receives considerable
contributions from strange quarks, one has in addition
$\mathrm{Im}(\delta^{d}_{22})_{LR} \lsim 10^{-5}
-10^{-6}$~\cite{phase:edm}. 
These bounds impose a stringent constraint on most
supersymmetric models, implying that the CP violating phases are 
very small and (or) that SUSY soft breaking terms must have a 
special flavour structure.

Small CP phases can be motivated by an approximate CP symmetry and
as previously pointed out, USY provides a natural scenario where all
Yukawa coupling phases are bound to be $\lsim 10^{-1}$.
In this work we take as a guideline the assumption that 
all supersymmetric phases should be no greater than the largest of the
phases present in the Yukawa textures of Eqs.~(\ref{usy:ex:2}). 

\begin{figure}[ht]
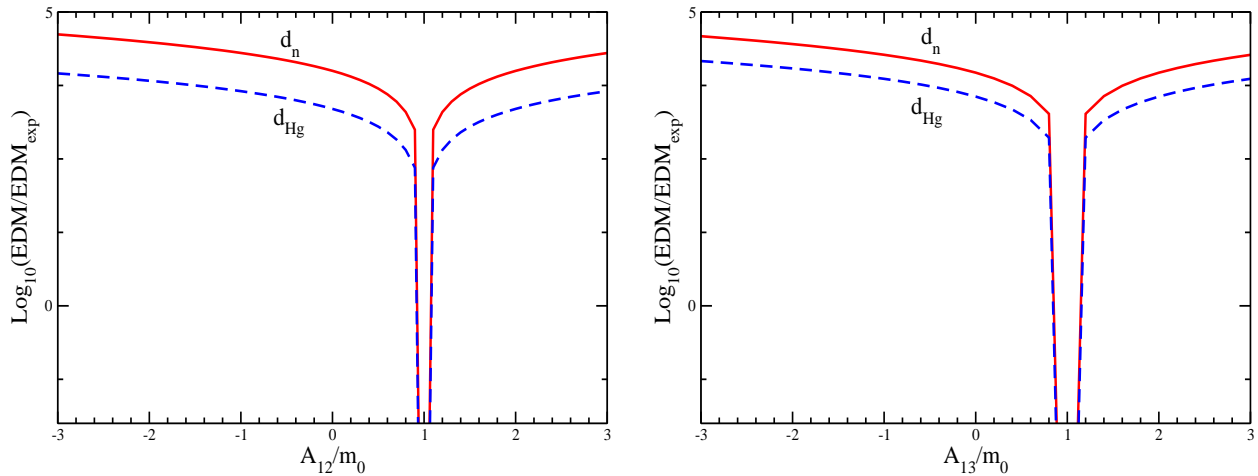

\begin{center}
\hspace*{-7mm}
\epsfig{file=na12.eps,width=8cm,height=6.25cm} \quad
\epsfig{file=na13.eps,width=8cm,height=6.25cm}\\
\caption{Neutron and mercury EDMs as function of the off--diagonal 
entries $A_{12}$ and
$A_{13}$ for $\tan \beta=5$ and $m_0=m_{1/2}=250$ GeV. 
$A_{ij}=1\;\forall ij \neq 12 (13)$ on the left (right) figure.}
\label{susy:edm:fig1}
\end{center}
\end{figure}

Although in many models the trilinear terms do not have the same
structure of the Yukawa couplings, when introducing the USY ansatz
in a supersymmetric framework the most appealing scenario would be
to have a USY texture in the $A$-terms as well. Nevertheless, we 
have verified that such textures for the $A$-terms do not suceed in
providing sizable SUSY contributions to the CPV observables without
violating the EDM bounds. Therefore we consider a more general scenario for the
trilinear terms, and allow for non-universalities in the magnitude
of the several elements.
We will show that, even in the limit of small (or vanishing) 
supersymmetric phases, the large mixing inherent to USY couplings
(displayed in $S_L$ and $S_R$), together with the experimental limits
on EDMs, severely constrain any non--universality of the trilinear terms.
In Fig.~\ref{susy:edm:fig1} we present the constraints from the 
EDMs on the off--diagonal entries of the $A$-terms, in particular 
$A_{12}$ and $A_{13}$. In our analysis we assumed $\tan \beta=5$,
$m_0= m_{1/2}= 250\;\GeV$ and 
$A_{ij} = m_0$ for all elements except $A_{12,13}$, which we set in
the range $[-3 m_0, 3 m_0]$.
 
\vspace*{8mm}

From these figures, it is clear that any significant 
non--universality among the $A$ terms (e.g. at $10\%$) 
leads to unacceptably large EDMs. Similar constraints hold for the other 
off--diagonal elements $A_{21}$, $A_{31}$, $A_{23}$ and $A_{32}$.
It is also worth noticing that such constraints are far more severe than those 
obtained in the case of hierarchical Yukawa couplings, as a consequence of the
already mentioned large intergenerational mixing.

As recently shown, this problem can be overcome if one takes 
the Yukawa couplings and the $A$--terms to be Hermitian, in order to ensure
that the diagonal elements of $A_{ij} Y_{ij}$ are real in any 
basis~\cite{khalil}. Another interesting possibility is to have trilinear
terms that can be factorized as $\hat{A} = A_{ij} Y_{ij} = A.Y$ or $Y.A$
~\cite{vives}. As above referred, this factorization implies that the 
mass insertion $(\delta_{11}^d)_{LR}$ is 
suppressed by the ratio $m_d/m_{\tilde{q}}$.
 
As an example for the trilinear couplings
that avoids the EDM constraints and implies signficant SUSY CP violation, we
consider the following structure:

\begin{equation}\label{Aterms}
A = m_0 \left(\begin{array}{ccc}
a & a & a \\
b & b & b \\
c & c & c \end{array} \right) \;.
\end{equation}
In this case the trilinear couplings $\hat{A}$ can be written as 
\begin{equation}
\hat{A} = \mathrm{diag}(a,b,c) \cdot Y \;.
\end{equation}
With this structure for the trilinear couplings, 
the EDMs do not impose any constraint on the parameters $a,b$, $c$ if these
are real, and the associated EDM values 
are similar to those obtained with universal and real $A$ terms. 
For complex $a,b$, and $c$, the constraints on the phases 
$\varphi_a$, $\varphi_b$, and $\varphi_c$ are shown in Fig. 2.\\ 
\begin{figure}[ht]
\begin{center}
\hspace*{-7mm}
\epsfig{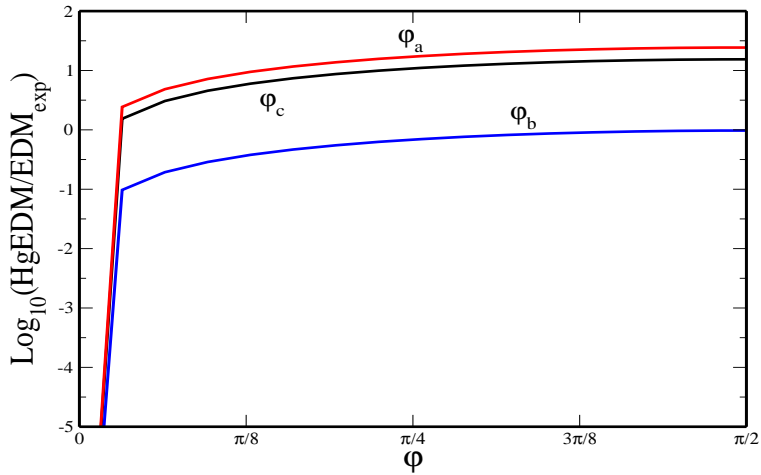}\\
\caption{The mercury EDM as function of the trilinear SUSY phases $\varphi$ 
for $\tan \beta=5$, $m_0=m_{1/2}=250$ GeV. $\vert A_{ij}\vert=1\;\forall ij$.}
\label{susy:edm:fig2}
\end{center}
\end{figure}
As it can be observed for the case of universal $|A_{ij}|$,
compatibility with experiment requires
\begin{equation}\label{susy:edm:dhgm}
\varphi_a \lsim 0.05 \; \mathrm{rad} \quad \; \quad \quad 
\varphi_{b} \lsim \pi/2 \;  \quad \;; \quad \quad 
\varphi_c \lsim 0.07 \; \mathrm{rad}\;.
\end{equation}
The reason why the phase $\varphi_b$ is much less constrained than
$\varphi_a$ and $\varphi_c$ is that the mixing between
the first and the second generations, as observed from 
Eqs.~(\ref{usy:ex:sul},\ref{usy:ex:sdl}), is smaller than the mixing 
between the first and the third generations. 
For $\vert A_{ij} \vert \simeq 3 m_0$, the EDMs constraint
on these phases becomes more stringent: 
\begin{equation}\label{susy:edm:dhgm0}
\varphi_a \lsim 0.02 \; \mathrm{rad} \quad \;, \quad \quad 
\varphi_{b} \lsim 0.35 \; \mathrm{rad}\quad \;, \quad \quad 
\varphi_c \lsim 0.02 \; \mathrm{rad}\;.
\end{equation}

Thus, we can conclude that in USY models the maximal allowed
non-universality for the $A$-terms is the structure presented in  
Eq.~(\ref{Aterms}), with the associated SUSY phases
($\varphi_{ij}$) within the limits given in Eq.~(\ref{susy:edm:dhgm0}).
In view of this, we analyze the contributions to the kaon
system CPV observables, $\varepsilon_K$ and $\varepsilon^\prime/\varepsilon$.

%
\subsection{\bf SUSY contributions to $\varepsilon_K$ and 
$\varepsilon^\prime/\varepsilon$}\label{susy:eK}

We proceed to investigate how CP violation can further constrain the
parameter space for the case of supersymmetric USY models.
In the kaon system, the most relevant observables are $\varepsilon_K$
and $\varepsilon^\prime/\varepsilon$. We first turn our attention to 
the indirect CP violation parameter, 
$\varepsilon_K$, which is given by
\begin{equation}\label{susy:eK:defek}
\varepsilon_K = \frac{e^{i \pi/4}}{\sqrt{2} \Delta m_K} \Im 
M_{12} \;,
\end{equation}
where $M_{12}$ is the off-diagonal entry in the kaon mass
matrix and $\Delta m_K$ is the mass difference of the short and
long-lived kaon states. 
The kaon mass matrix element can be defined as follows
\begin{equation}\label{susy:eK:defm12}
2 m_K M_{12}^* =
\av{\bar{K}^0|\mathcal{H}_{\mathrm{eff}}^{\Delta S=2} |K^0}\;,
\end{equation}
where $\mathcal{H}^{\Delta S=2}_{eff}$ is the effective Hamiltonian (EH) 
for $\Delta S = 2$ 
transitions, 
which can be expressed in the operator product expansion as
\begin{equation}\label{susy:eK:Hs2eff}
\mathcal{H}^{\Delta S=2}_{eff} = \sum_i C_i (\mu) Q_i\;.
\end{equation}
In the above formula, $C_i (\mu)$ are the Wilson coefficients 
and $Q_i$ are the EH local operators.\\
The relevant operators for the gluino contribution are~\cite{EH}
\bea\label{susy:eK:oper}
Q_1 & = & \bar{d}^{\alpha}_L \gamma_{\mu} s^{\alpha}_L \bar{d}^{\beta}_L 
\gamma^{\mu} s^{\beta}_L,~~~~~~
Q_2 = \bar{d}^{\alpha}_R s^{\alpha}_L \bar{d}^{\beta}_R 
s^{\beta}_L,~~~~~~~
Q_3  =  \bar{d}^{\alpha}_R s^{\beta}_L \bar{d}^{\beta}_R 
s^{\alpha}_L,\nonumber\\
Q_4 &=&  \bar{d}^{\alpha}_R s^{\alpha}_L \bar{d}^{\beta}_L 
s^{\beta}_R,~~~~~~~~~~~~
Q_5 =  \bar{d}^{\alpha}_R s^{\beta}_L \bar{d}^{\beta}_L 
s^{\alpha}_R,
\eea
as well as the operators $\tilde{Q}_{1,2,3}$, that are 
obtained from $Q_{1,2,3}$ by the exchange $L \leftrightarrow R$. 
In the latter equations, $\alpha$ and $\beta$ are $SU(3)_c$ colour
indices, and the colour matrices obey the normalization 
$\Tr(t^a t^b)=\delta^{ab}/2$. Due to the gaugino dominance 
in the chargino--squark loop,
the most significant $\tilde{\chi}^\pm$ contribution is associated with the 
operator $Q_1$~\cite{olegchar}. 

In the presence of supersymmetric ($\tilde{g}$ and $\tilde{\chi}^\pm$) 
contributions, the 
following result for the amplitude $M_{12}$ is obtained:
\begin{equation}\label{susy:eK:fullm12}
M_{12} = M_{12}^{\mathrm{SM}} +  
M_{12}^{\tilde{g}} +  M_{12}^{\tilde{\chi}^\pm}\;.
\end{equation}
The SM contribution can be written as~\cite{buras:2001,branco:book}
\begin{equation}\label{susy:eK:m12sm}
M_{12}^{\mathrm{SM}} = \frac{G_F^2 m^2_W}{12 \pi^2}
f^2_K m_K \hat{B}_K \mathcal{F}^* \;,
\end{equation}
where $f_K$ is kaon decay constant, $\hat{B}_K$ is the bag parameter, 
and the function $\mathcal{F}$ can be found in ~\cite{branco:book}.
For the specific USY ansatz we are considering, we find that 
the SM contribution to $\varepsilon_K$ is  
$\mathcal{O}(10^{-5})$.
The supersymmetric term $M_{12}^{\tilde{g}}$ is given by
\bea\label{susy:eK:m12gluino}
M_{12}^{\tilde{g}} &=& \frac{-\alpha_S}{216 m_{\tilde{q}^2}} \frac{1}{3}
m_{K}f_{K}^{2}
\Biggl\{  \left( (\delta^d_{12})^2_{LL} +(\delta^d_{12})^2_{RR} \right) 
        B_1(\mu) \left(  24\,x\,f_6(x) + 66\,\tilde{f}_6(x) \right)
        \nonumber \\
&+&(\delta^d_{12})_{LL} (\delta^d_{12})_{RR} \left(\frac{m_{K}}{m_{s}
(\mu)+m_{d}(\mu)}\right)^{2}
\biggl[B_4(\mu) \left(378 \,x\,f_6(x) -54 \tilde{f}_6(x)\right) \nonumber\\
&+& B_5(\mu) \left(6 \,x\,f_6(x) +30 \tilde{f}_6(x)\right) \biggr]
+\left((\delta^d_{12})^2_{LR} + (\delta^d_{12})^2_{RL} \right)
\left(\frac{m_{K}}{m_{s}(\mu)+m_{d}(\mu)}\right)^{2}\nonumber\\
&&\left(-\frac{255}{2} 
B_2(\mu) -\frac{9}{2} B_3(\mu) \right) x\, f_6(x) +
(\delta^d_{12})_{LR} 
(\delta^d_{12})_{RL} \left(\frac{m_{K}}{m_{s}(\mu)+m_{d}(\mu)}
\right)^{2}\nonumber\\
&&\biggl[-99 B_4(\mu) - 45 B_5(\mu) \biggr]\tilde{f}_6(x)
\Biggr\},
\end{eqnarray}
where $x=(m_{\tilde{g}}/m_{\tilde{q}})^2$ and 
the functions $f_6(x),
\tilde{f}_6(x)$ are given in Ref.\cite{EH}. 
We have used the matrix elements of the 
normalized operators $Q_i(\mu)$ with the following $B$-parameters at 
$\mu = 2 \GeV$~\cite{EH}
\bea\label{susy:eK:Bpar}
B_1(\mu) &=& 0.60\;, ~~~~~~~~~~ B_2(\mu) = 0.66\;, ~~~~~~~~~~ 
B_3(\mu) = 1.05\;,\nonumber\\
B_4(\mu) &=& 1.03\;,~~~~~~~~~~ B_5(\mu) = 0.73\;. \nonumber
\eea
Finally $ M_{12}^{\tilde{\chi}^\pm}$ is~\cite{olegchar}
\begin{equation}\label{susy:eK:m12char}
M_{12}^{\tilde{\chi}^\pm} = \frac{g^2}{768 \pi^2 m_{\tilde{q}^2}} \frac{1}{3}
m_{K}f_{K}^{2} B_1(\mu) \left(\sum_{a,b} K_{a2}^* 
(\delta^u_{LL})_{ab} K_{b1}\right)^2
\sum_{i,j} \vert V_{i1}\vert^2 \vert V_{j1} \vert^2 H(x_i,x_j),
\end{equation}
where $x_i=(m_{\tilde{\chi}_i^+}/m_{\tilde{q}})^2$. 
Here $K$ is the CKM matrix, 
$a,b$ are flavour indices, $i,j$ label the 
chargino mass eigenstates, and $V$ is one of the matrices used for
diagonalizing the chargino mass matrix.
The loop function $H(x_i, x_j)$ is given in Ref.\cite{olegchar}.

\begin{figure}[ht]
\begin{center}
\hspace*{-7mm}
\epsfig{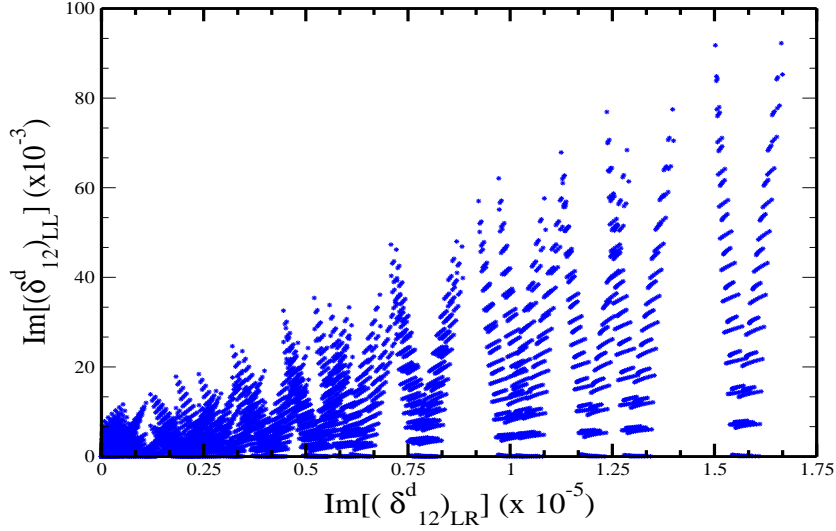} \\
\caption{Correlation between $(\delta_{12}^d)_{LL}$ and 
$(\delta_{12}^d)_{LR}$ mass inserions for for $\tan \beta=5$, 
$m_0=m_{1/2}=250$ GeV. $a,b,c$ vary from $[-3,3]$ and $\varphi_{a,b,c}$ 
are chosen so that the EDMs are consistent with the experimental limits.}
\label{susy:edm:fig3}
\end{center}
\end{figure}

To saturate $\varepsilon_K$ from gluino contributions,
one should have~\cite{masiero}
\begin{equation}\label{susy:eK:deltalim}
\sqrt{\vert \mathrm{Im}(\delta_{12}^d)_{LL}^2 \vert} \sim 10^{-3}
\quad \mathrm{or} \quad
\sqrt{\vert \mathrm{Im}(\delta_{12}^d)_{LR}^2 \vert} \sim 10^{-4}\;,
\end{equation}
whilst the chargino contributions require
\begin{equation}\label{susy:eK:chardeltalim}
\mathrm{Im}(\delta_{12}^u)_{LL}^2 \sim 10^{-1}\;.
\end{equation}
In the present scenario, 
using the USY texture of Eq.~(\ref{usy:ex:2}) and $A$-terms 
as in Eq.~(\ref{Aterms}), we find the correlation between 
these two mass insertions is as given in Fig.~\ref{susy:edm:fig3}.\\
\vspace*{3mm}

From Figure~\ref{susy:edm:fig3}, one can see
that the $LL$ mass insertions give the dominant 
gluino contribution to saturate $\varepsilon_K$, contrary to the $LR$ mass 
insertions, which are typically much smaller than the required value
given in Eq.~(\ref{susy:eK:deltalim}). In addition, we also 
find (for $m_0=m_{1/2}=250\;\GeV$) that 
$\mathrm{Im}(\delta^u_{12})_{LL} \simeq 10^{-3}$,
two orders of magnitude below the required value, 
so that one has a negligible chargino contribution  to $\varepsilon_K$. 
Therefore, with USY Yukawa couplings, the MSSM with non--universal $A$-terms
(as those given in Eq.~(\ref{Aterms})) can account for the experimentally 
observed indirect CP violation in the kaon system. 

It is important to note that the behaviour of the $LL$ mass insertion
(particularly the significant dependence on the non--universality of 
the $A$--terms) is in clear contrast with the case of standard hierarchical 
Yukawa couplings. It is well known 
that in these scenarios, and with universal scalar soft masses, 
$\mathrm{Im}(\delta^d_{12})_{LL}$ is typically of order $10^{-5}$ and 
the $LL$ mass insertions are independent of the $A$--terms. 
In our model, the large mixing displayed by the rotation
matrices enhances any non-diagonal contribution to $M^2_{\tilde{Q}_L}$
induced by the trilinear terms from RG evolution, hence $LL$
mass insertions can account for the experimental value of $\varepsilon_K$.

Next we consider SUSY contributions to  
$\varepsilon^\prime/\varepsilon$. The grand average of the
experimental values is~\cite{fanti:1999,NA48:0101},
\begin{equation}\label{susy:eK:eprimeexp}
\Re(\varepsilon^\prime/\varepsilon)=(19.2 \pm 2.4)\times10^{-4}\;.
\end{equation}
$\varepsilon^{\prime}$ can be given as
ratio of isospin amplitudes $A_I$, which are defined as 
\begin{equation}\label{susy:eK:ai}
A_I e^{i \delta_i}=
\langle \pi \pi(I) \vert \mathcal{H}_{eff}^{\Delta S=1} \vert K^0 \rangle\;,
\end{equation} 
where $I=0,2$ is the isospin of the final two-pion state.
Therefore we can write~\cite{branco:book}
\begin{equation}\label{susy:eK:epA}
\frac{\varepsilon^\prime}{\varepsilon} 
\sim \frac{i}{\sqrt{2}} \;e^{i(\delta_2-\delta_0)}\;
\frac{\Im(A_2 A_0^*)}{{|\varepsilon A_0|}^2}\;.
\end{equation}

Assuming that the strong phase difference ($\delta_2-\delta_0$) 
is close to $\pi/4$,
the expression for $\varepsilon^\prime$ can be rewritten as
\begin{equation}\label{susy:eK:epomega}
\frac{\varepsilon^\prime}{\varepsilon} = e^{i \frac{\pi}{4}} 
\frac{\omega}{\sqrt{2} |\varepsilon|} \xi (\Omega -1)
\end{equation}
with 
\begin{equation}\label{susy:eK:epomegadef}
\omega= \mathrm{Re} A_2/\mathrm{Re} A_0\;, \quad 
\xi =\frac{\Im A_0}{\Re A_0}\quad \mathrm{and} \quad
\Omega = \frac{\Im A_2}{\omega \Im A_0}\;.
\end{equation}
The real parts of both amplitudes are experimentally known to be 
$\Re A_2 = 1.5 \times 10^{-8}\;\GeV$ and 
$\Re A_0 = 33.3 \times 10^{-8}\;\GeV$.

The effective Hamiltonian for $\Delta S=1$
transitions receives contributions from operators with
dimension four, five and six, and can be symbolically written as 
\begin{equation}\label{susy:eK:Heff}
\mathcal{H}_{\mathrm{eff}}^{\Delta S=1} 
= \mathcal{H}_{\mathrm{eff}}^{(4)}
+ \mathcal{H}_{\mathrm{eff}}^{(5)}
+ \mathcal{H}_{\mathrm{eff}}^{(6)}\;.
\end{equation}
The $\Delta S=1$ effective Hamiltonian containing 
dimension-6 operators is given by~\cite{jamin:1993} 
\begin{equation}\label{susy:eK:epHeff6}
\mathcal{H}_{\mathrm{eff}}^{(6)} 
= \sum_{i=1}^{10 } C_i(\mu) O_i +\hc \;,
\end{equation}
where $C_i$ are the Wilson coefficients. $O_{1,2}$ refers to the
current--current operators, $O_{3-6}$ to the QCD penguin operators and 
$O_{7-10}$ to the electroweak penguin operators. In addition, one
should take into account $\tilde{O}_i$ operators, which 
are simply related to $O_i$ by the exchange $L \leftrightarrow R$. 
The corresponding matrix elements  $\langle \pi \pi \vert 
\tilde{O}_i \vert K^0 \rangle$ can be 
obtained from the matrix elements of $\langle \pi \pi \vert 
O_i \vert K^0 \rangle$, multiplying them by $(-1)$ while the 
$\tilde{C}_i$ are obtained from $C_i$ by exchange $L \leftrightarrow R$.
In principle, there are also two
dimension-5 ``magnetic'' operators $O_{\gamma}$ and $O_g$~\cite{fabbrichesi}, 
which are induced by gluino exchanges:
\begin{equation}\label{susy:eK:epHeff5}
\mathcal{H}_{\mathrm{eff}}^{(5)} 
= C_{\gamma} O_{\gamma} + C_{g} O_g + \hc \;,
\end{equation}
where 
\begin{eqnarray}\label{susy:eK:epCHeff5}
C_{\gamma} &=& \frac{\alpha_s \pi}{m_{\tilde{q}}^2} \left[
(\delta^d_{12})_{LL} \frac{8}{3} M_3(x_{gq})+ 
(\delta^d_{12})_{LR} \frac{m_{\tilde{g}}}{m_s} \frac{8}{3} M_1(x_{gq})
\right] \;, \nonumber\\
C_{g} &=& \frac{\alpha_s \pi}{m_{\tilde{q}}^2} \left[
(\delta^d_{12})_{LL} 
\left(-\frac{1}{3} M_3(x_{gq}) - 3 M_4(x_{gq}) \right) +
(\delta^d_{12})_{LR} \frac{m_{\tilde{g}}}{m_s} 
\left(-\frac{1}{3} M_1(x_{gq}) - 3 M_2(x_{gq}) \right)
\right] \;.\nonumber\\
\end{eqnarray}
Here $x_{gq}=m_{\tilde{g}}^2/m_{\tilde q}^2$ and the 
loop functions $M_1$, $M_2$, $M_3$ and $M_4$ are given in Ref.~\cite{masiero}. 
The operators can be written as:
\begin{eqnarray}\label{susy:eK:epOHeff5}
O_{\gamma} &=& \frac{Q_d~e}{8\pi^2} m_s \bar{d}^{\alpha}_L \sigma^{\mu \nu}
s^{\alpha}_R F_{\mu \nu}\;,\nonumber \\
O_{g} &=& \frac{g}{8\pi^2} m_s \bar{d}^{\alpha}_L \sigma^{\mu \nu}
t^A_{\alpha \beta} s^{\beta}_R G_{\mu \nu}^A \;.
\end{eqnarray}
One should also consider the $\tilde{C}_{\gamma,g}\;
\tilde{O}_{\gamma,g}$ contributions to
$\mathcal{H}_{\mathrm{eff}}^{(5)}$, which are computed from
Eq.~(\ref{susy:eK:epOHeff5}) as above described.

The only dimension-four contribution relevant for this calculation is
associated with the operator $Q_Z$, generated by the $\bar{s}dZ$
vertex which is mediated by chargino exchanges. 
\begin{equation}\label{susy:eK:epHeff4} 
\mathcal{H}_{\mathrm{eff}}^{(4)} =
 - \frac{G_F}{\sqrt{2}} \frac{e}{\pi^2} m_Z^2 \tan \theta_W
Z_{ds} \bar{s}_L \gamma_{\mu} Z^{\mu} d_L + \hc\;.
\end{equation}
The coefficient $Z_{ds}$ is given by~\cite{isidori}
\begin{equation}
Z_{ds}= \lambda_t C_0(x_t) + \bar{\lambda}_t H_0(x_{q\chi}) \;,
\end{equation}
where the first term refers to the SM $Z^0$ penguin 
contribution, evaluated in the 't~Hooft--Feynman gauge, 
and the second one represents the SUSY contribution associated with 
$\bar{\lambda}_t= (\delta^u_{23})_{LR}
(\delta^u_{13})^*_{LR}$.
Finally $x_{t}= m_{t}^2/m_{W}^2$, 
$x_{q\chi}= m_{\tilde{q}}^2/m_{\chi^{\pm}}^2$ 
and the loop functions are given by~\cite{isidori}
\begin{equation}\label{susy:eK:C0H0}
C_0(x)=\frac{x}{8} \left[\frac{x-6}{x-1} + \frac{3x+2}{{(x-1)}^2} 
\ln(x)\right]\;,\quad
H_0(x)= \frac{-x (x^3 -6 x^2 +3 x + 2 + 6 x \ln x )}{48 (1-x)^4}\;.
\end{equation} 
In USY scenarios, the SM prediction for $\varepsilon^\prime/\varepsilon$, 
which receives contributions from the operators $O_{6}$ and $O_{8}$,
as well as from $Z_{ds}$, is found to be $\mathcal{O}(10^{-6})$.
The supersymmetric contributions to $\varepsilon^\prime/\varepsilon$
could be dominated either by gluino or chargino exchanges. We decompose 
these contributions as follows:
\begin{equation}
\mathrm{Re}\left(\frac{\varepsilon^\prime}{\varepsilon}\right)^
{\mathrm{SUSY}} = \mathrm{Re}\left(\frac{\varepsilon^\prime}{\varepsilon}
\right)^{\tilde{g}}+ \mathrm{Re}\left(\frac{\varepsilon^\prime}{\varepsilon}
\right)^{\tilde{\chi}^\pm} .
\end{equation} 
The first term is associated with gluino mediated diagrams and receives
contributions from the ``magnetic'' operators in
Eq.~(\ref{susy:eK:epOHeff5}). 
The matrix element $\langle {(\pi \pi)}_{0,2}|O_\gamma|K^0 \rangle$ is
proportional to the photon condensate, hence it must be very small. 
Accordingly, we shall neglect the $C_{\gamma}$
contributions to $\varepsilon^\prime/\varepsilon$.
The matrix elements of the operator $O_g$ are given by~\cite{fabbrichesi}:
\begin{eqnarray}
\av{{(\pi \pi)}_{2}|O_g|K^0}&=&0\;,\nonumber \\
\av{{(\pi \pi)}_{0}|O_g|K^0}&=&
\sqrt{\frac{3}{2}} \av{{(\pi^+  \pi^-)}|O_g|K^0}=\nonumber \\
&=&-\sqrt{\frac{3}{2}} \frac{1}{16 \pi^2} \frac{11}{2} 
\frac{m_s}{m_s+m_d} \frac{f^2_K}{f^3_{\pi}} m^2_K m^2_{\pi} B_g^{(1/2)}\;,
\end{eqnarray}
where $B_g^{(1/2)} \sim 1$ and $f_{\pi}$ is the pion decay constant. 
Thus, the chromomagnetic 
contribution to $\varepsilon^\prime/\varepsilon$ is 
\begin{equation}\label{epsprimeG}
\mathrm{Re}\left(\frac{\varepsilon^\prime}{\varepsilon}
\right)^{\tilde{g}} \simeq 
\frac{11 \sqrt{3}}{64 \pi^2 \vert \varepsilon \vert 
\mathrm{Re} A_0}~  \frac{m_s}{m_s+m_d} \frac{f_K^2}{f_{\pi}^3}~ m_K^2~
m_{\pi}^2~ \mathrm{Im}\left[ C_g - \tilde{C}_g\right] \;.
\end{equation}

We now turn our attention to the chargino contributions. In addition
to terms proportional to a single mass insertion
$(\delta^u_{ab})_{LL}$~\cite{olegchar}, one has to take into account 
other contributions involving a double mass insertion, like those
arising from the supersymmetric effective $\bar{s}dZ$ vertex. 
The term $(\varepsilon^\prime/\varepsilon)^{\tilde{\chi}^\pm}$ can be
separated into
\begin{equation}\label{susy:eK:epchar}
\mathrm{Re}\left(\frac{\varepsilon^\prime}{\varepsilon}\right)^
{\tilde{\chi}^\pm} = \mathrm{Re}\left(\frac{\varepsilon^\prime}{\varepsilon}
\right)^{\tilde{\chi}^\pm}_{1\mathrm{MIA}} + 
\mathrm{Re}\left(\frac{\varepsilon^\prime}{\varepsilon}
\right)^{\tilde{\chi}^\pm}_{2\mathrm{MIA}}.
\end{equation}
As in Ref.~\cite{olegchar}, the first term is given by
\begin{equation}\label{susy:eK:epchar2}
{\left(\frac{\varepsilon^\prime}{\varepsilon}\right)}^{\tilde{\chi}^\pm} 
_{1\mathrm{MIA}}=
\Im \left(\sum_{a,b} K_{a2}^* 
(\delta^u_{ab})_{LL} K_{b1}\right) \; F_{\varepsilon^\prime}({x_{q\chi}})\;.
\end{equation}
As before $K$ denotes the CKM matrix, $a,b$ are flavour indices and the
function $F_{\varepsilon^\prime}(x_{q\chi})$, 
with $x_{q\chi}=m_{\tilde{\chi}^\pm}^2/m_{\tilde q}^2$, 
is given in Ref.\cite{olegchar}. This
contribution is clearly dominated by $(\delta^u_{12})_{LL}$.
The second term on the r.h.s. of Eq.~(\ref{susy:eK:epchar}) can be
written as 
\begin{equation}\label{epsprimeZ}
\mathrm{Re}\left(\frac{\varepsilon^\prime}{\varepsilon}
\right)^{\tilde{\chi}^\pm}_{2\mathrm{MIA}}= 
\left[ 1.2 - R_s~ \vert r_Z^{(8)} \vert 
~B_8^{(3/2)} \right]~ \mathrm{Im}\Lambda_t \;, 
\end{equation}
where the parameters $R_s$ and $r_Z^{(8)}$ are given in Ref.~\cite{isidori}.
$B^{(3/2)}_8$ is the non--perturbative parameter describing the hadronic 
matrix element and $\Lambda_t$ is given as 
\begin{equation}
\Lambda_t =\left[ (\delta^u_{23})_{LR} (\delta^u_{13})^*_{LR} \right]
~H_0(x_{q\chi})\;, 
\end{equation}
where $H_0$ has been defined in Eq.~(\ref{susy:eK:C0H0}).

Now let us discuss in particular the r\^ole of each contribution within
our model.
As pointed out, the gluino contributions to $\varepsilon^\prime/\varepsilon$ 
can naturally saturate the experimental values if the trilinear couplings are 
non--universal, since the required value of the $LR$ mass insertion is 
comparatively weaker, $(\mathrm{Im}(\delta^d_{12})_{LR} \sim 10^{-5})$.
As mentioned above, it is possible to obtain such values and still
saturate $\varepsilon_K$, so the chromomagnetic operator $O_g$ can
give a significant contribution to  $\varepsilon^\prime/\varepsilon$. 
Furthermore, and as aforementioned, in USY models the $LL$ mass
insertions are also enhanced, hence one might expect significant 
contributions from the chargino as well.
 
\begin{figure}[ht]
\begin{center}
\epsfig{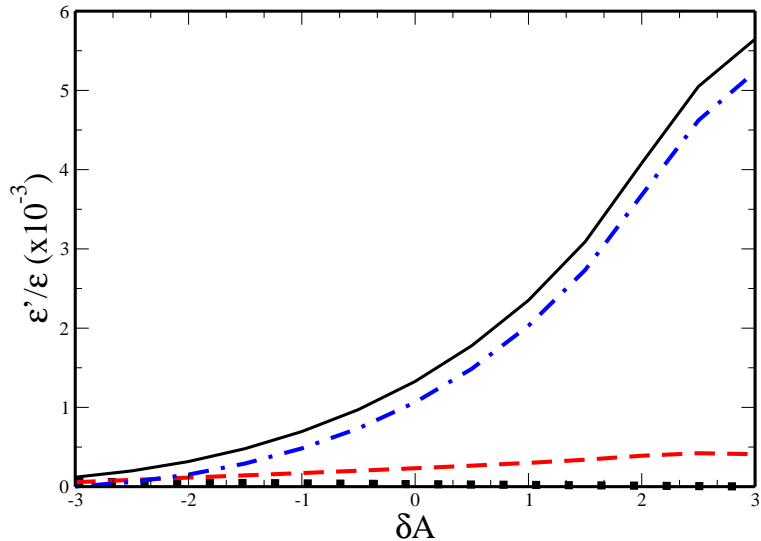}\\
\caption{Contributions to $\varepsilon^\prime/\varepsilon$ versus
the non--universality parameter $\delta A$.
The dotted line refers to gluino contributions, the
dott-dashed and dashed lines to the chargino one 
(with one and two mass insertions, respectively). The solid
line stands for the total supersymmetric contribution to  $\varepsilon^\prime/\varepsilon$.}
\label{susy:epsprime:phib}
\end{center}
\end{figure}

In Fig.~\ref{susy:epsprime:phib} we plot the gluino and 
chargino contributions to 
$\varepsilon^\prime/\varepsilon$ as function of the 
parameter $\delta A$ that parametrises the non--universality of the 
trilinear couplings.
\begin{equation} 
\delta A = (A_{3i} - A_{1i,2i})/m_0 = c-a \quad (\mathrm{assuming}\;a=b) \;.
\end{equation}
The SUSY CP violating phases have been fixed as follows: 
$\varphi_a=\varphi_c=0$ and $\varphi_b=0.1$. 
As in the previous figures, we have assumed $\tan \beta=5$ and 
$m_0=m_{1/2}=250$ GeV. 

As can be seen from this figure, the chargino contributions with one mass 
insertion give the dominant contribution to $\varepsilon^\prime/\varepsilon$.
The chargino contributions with two mass insertions that arise from the 
SUSY effective $\bar{s}dZ$ vertex are also relevant and considerably
larger than that of the gluino.

%
\section{\bf \large Small CP phases versus large CP asymmetry \pmb{$
a_{J/\psi K_S}$}}
\label{bp}

For a long time all experimental evidence of CP violation was confined to 
the kaon sector. However, recent results from the $B$--factories have
confirmed the existence of CP violation in the $B$ sector as well.
The fact that the asymmetry $a_{J/\psi K_S}$ is large may lead to 
the conclusion that CP cannot be viewed as an
approximate symmetry. We will show next that this is not necessarily true. 
In fact, it might happen that models with small CP
violating phases but with a considerably large flavour mixing can also
account for the recent experimental results.
As it was shown in previous sections, supersymmetric USY models
are indeed an example of this class of models, where a large flavour
mixing and small CP phases still succeed in saturating the measured values
of the CPV observables.
In this section we address some of the implications of supersymmetric
USY models for the $B$--system. 

Regarding the CP asymmetry of the $B_d$ and $\bar{B}_d$ 
meson decay to the CP eigenstate $J/\psi K_S$, defined by
\begin{equation}
a_{J/\psi K_S}(t) = \frac{\Gamma(B_d(t) \to \psi K_S) - 
\Gamma(\bar{B}_d(t) \to \psi K_S)}{\Gamma(B_d(t) \to \psi K_S) +
\Gamma(\bar{B}_d(t) \to \psi K_S)}= - a_{J/\psi K_S} \sin(\Delta
m_{B_d} t)\;,
\end{equation}
BaBar and Belle give the following results
\begin{align}
a_{J/\psi K_S}&= 0.741 \pm 0.067 \pm 0.034 \quad \quad 
\mathrm{(BaBar)} \;, \nonumber \\
a_{J/\psi K_S}&= 0.719 \pm 0.074 \pm 0.035 \quad \quad
\mathrm{(Belle)} \;,
\end{align}
which dominate the present world average, 
$a_{J/\psi K_S}=0.734 \pm 0.054$~\cite{2beta:babar,2beta:belle,2beta:2001}.

In the case of the Standard Model, $a_{J/\psi K_S}$
can be easily related to one of the inner angles of the unitarity
triangle:
\begin{equation}
a_{J/\psi K_S}^{SM}=\sin 2 \beta\;; \quad
\beta \equiv \arg \left(- \frac{{V}_{cd} {V}_{cb}^* }
{{V}_{td} {V}_{tb}^*} \right) \;.
\end{equation}
In the USY ansatz of Eq.~(\ref{usy:ex:2}), SM contributions to 
$\sin 2 \beta$ are negligible, $\mathcal{O}(10^{-3})$.
In view of this, it is mandatory that new physics provides a large contribution
to $B_d-\bar{B}_d$.

In the case of supersymmetric contributions to
the $\Delta B= 2$ transition, the off--diagonal element of the $B_d$ mass
matrix can be written as 
\begin{equation}
M_{12}(B_d) = 
\frac{\langle B_d \vert \mathcal{H}_{\mathrm{eff}}^{\Delta B=2} 
\vert \bar{B}_d 
\rangle}{2 m_{B_d}} = M_{12}^{\mathrm{SM}}(B_d) + M_{12}^{\mathrm{SUSY}}(B_d).
\end{equation}
The supersymmetric effects are usually parametrised through
\begin{equation}
r_d^2 e^{2 i \theta_d} = \frac{M_{12}(B_d)}{M_{12}^{\mathrm{SM}}(B_d)},
\end{equation}
where $\Delta m_{B_d} = 2 \vert M_{12}^{\mathrm{SM}}(B_d)\vert r_d^2$. 
Thus, in the presence of SUSY contributions, the 
CP asymmetry parameter is now given by
\begin{equation}
a_{J/\psi K_S} = \sin (2 \beta + 2 \theta_d)\;.
\end{equation}
Hence, the measurement
of $a_{J/\psi K_S}$ would not determine $\sin 2 \beta$ 
but rather $\sin(2 \beta + 2 \theta_d)$. 

\begin{figure}[ht]
\begin{center}
\epsfig{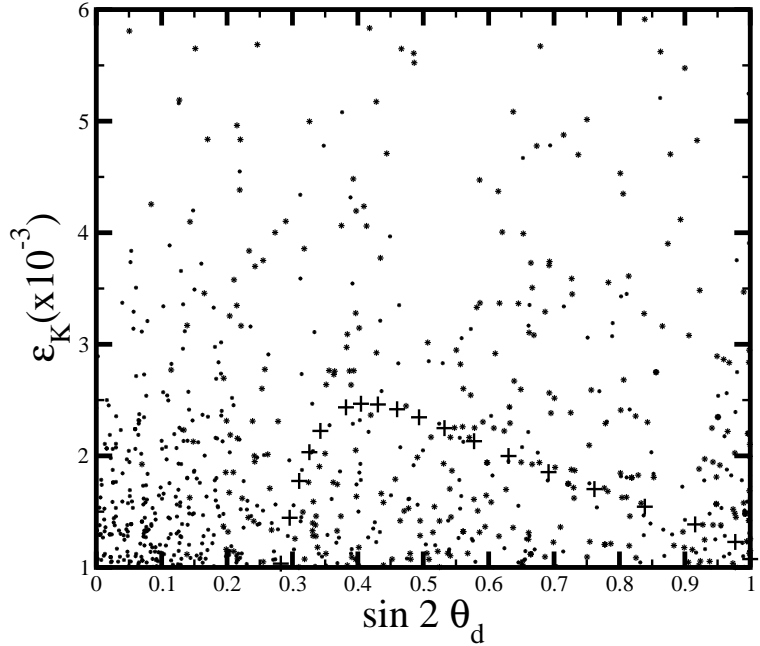}\\
\caption{Correlation between $\varepsilon_K$  and $\sin 2 \theta_d$.
The crosses (+) correspond to values 
$a=b=-3$ and $c \in \left[-1.5,0.5\right]$.}
\label{eps-beta}
\end{center}
\end{figure}

In many models, the SUSY contribution can be 
sufficient to account for the observed values of $a_{J/\psi K_S}$. 
However, this requires a large mixing in the $LL$ and $LR$ sectors, 
and this is indeed what we have in our USY example. As emphasized above, 
due to the large mixing in the USY textures, the non--universal 
$A$--terms have a non-negligible effect in the running of the squark masses
and although we imposed universality of the scalar soft masses at the
GUT scale, one finds considerable off diagonal contributions at the
electroweak scale, particularly associated with mixtures of the third
generation with the first two.
Since in this class of models flavour mixing is not supressed, one finds
that the $(\delta^{d,u}_{13})_{LL}$ and $(\delta^{d,u}_{13})_{LR}$ mass
insertions (the most relevant in the $a_{J/\psi K_S}$ calculation) can
be significantly large and saturate the experimental result as we show 
in detail below.
In our numerical analysis, we took into account charged--Higgs,
chargino and gluino contributions to 
$M_{12}^{\mathrm{SUSY}}(B_d)$, as given in Ref.~\cite{bertolini}.
The leading supersymmetric contribution 
to $\sin(2 \beta + 2 \theta_d)$ is associated with the 
$\Delta B=2$ chargino mediated box diagrams.
Instead of using the mass insertion approximation, as done in the
previous sections, we choose to compute the full contibutions to 
$M_{12}^{\tilde{\chi}^\pm}(B_d)$.  
The off--diagonal element of the $B_d$ mass matrix is accordingly
given by:
\begin{eqnarray}\label{susy:bb:charM12}
M_{12}^{\tilde{\chi}^\pm}(B_d) &=& \frac{\alpha_W^2}{16}
\frac{2}{3} f_B^2 B_{B_d} \eta^{-6/23} m_{B_d}
\sum_{h,k=1}^{6} \sum_{i,j=1}^{2} \frac{1}{m_{\tilde{\chi_j}^\pm}^2} 
\left\{ 
\left[
V^*_{j1}\; \Gamma_{U_L}^{kb} - 
V^*_{j2}\; {(\Gamma_{U_R}\;\hat{Y}_U\; K)}_{kb} \right] \right.
\nonumber \\
&& \left[
V_{i1}\; {\Gamma^{U_L}}^*_{kd} - 
V_{i2}\; {(\Gamma^{U_R}\;\hat{Y}_U\; K)}^*_{kd} \right]
\left[V^*_{j1}\; \Gamma^{U_L}_{hb} - 
V^*_{j2}\; {(\Gamma_{U_R}\;\hat{Y}_U\; K)}_{hb} \right] 
\nonumber \\
&& \left[
V_{j1}\; {\Gamma^{U_L}}^*_{hd} - 
V_{j2}\; {(\Gamma^{U_R}\;\hat{Y}_U\; K)}^*_{hd} \right]
G^\prime (x_{\tilde{u}_k \tilde{\chi_j}},
x_{\tilde{u}_h \tilde{\chi_j}},x_{\tilde{\chi_i} \tilde{\chi_j}} )
\left. \right\} \;, 
\end{eqnarray}
where $x_{ab} = (m_{a}/m_{b})^2$, $f_{B_d}$ is the $B_d$ meson 
decay constant, $B_{B_d}$ is the vacuum saturation parameter, 
and $\eta$ is the QCD correction factor.
As before, $V$ is one of the unitary matrices that diagonalise the
chargino mass matrix and $K$ is the CKM matrix. 
$\hat{Y_U} \equiv \diag(m_u, m_c, m_t)/(\sqrt{2} m_W \sin \beta)$,
$\Gamma_{U_{L,R}}$ are the matrices relating the squark interaction
and mass eigenstates, and the loop function $G^\prime$ is given in
Ref.~\cite{bertolini}.  
In Fig.~\ref{eps-beta} we present the correlation between the values of
$\varepsilon_K$ and $\sin 2 \theta_d$ for $\tan \beta =5$ and $m_0=m_{1/2}=
250$~GeV. The values of the parameters $a,b,$ and $c$ are randomly selected
in the range $[-3,3]$ and the phases are fixed as $\varphi_{a}=\varphi_{c}=0$
and $\varphi_b=0.1$.

\vspace*{12mm}

From Fig.~\ref{eps-beta}, 
it is clear that one can have $\sin 2 \theta_d$ within
the experimental range 
while having a prediction for $\varepsilon_K$ and
$\varepsilon'/\varepsilon$ compatible with the measured value. 
This result appears as a characteristic of the 
kind of models under consideration. Despite the smallness of the phases 
introduced in these models via the Yukawa and trilinear couplings, 
they are sufficient to account for CP
violation in the $K$-system, as well as the observed value of 
$a_{J/\psi K_S}$, which requires a phase of order one.
 
It is worth mentioning that the 
non--universality of the $A$-terms is crucial for enhancing the SUSY 
contributions. As emphasized above, for universal $A$-terms the mass 
insertions $(\delta^u_{31})_{LL}$ and $(\delta^u_{13})_{LR}$ are, 
respectively, of order $10^{-5}$ and $10^{-6}$, leading to a negligible 
$\sin 2 \theta_d$. When non--universal $A$-terms are considered, we find 
that the chargino contribution displays a large phase, while the 
gluino and SM contributions are almost real and in general smaller 
than the chargino contribution. The Higgs and neutralino contributions 
are negligible in the cases under consideration. 

If we assume that the universality of the $A$-terms 
is broken by setting $a=b\neq c$ as in Eq.~(\ref{Aterms}), it is 
possible to find sizable values for the the SUSY 
contributions to $a_{J/\psi K_S}$. In Fig.~\ref{eps-beta} we
considered particular values of $a,\;b,\;c$, namely
$a=b=-3$ and $c \in \left[-1.5,0.5\right]$, and marked the associated
contributions to $\sin 2\theta_d$ and $\varepsilon_K$ with a cross
($+$). As seen from Fig.~\ref{eps-beta}, for this set of points, 
$\sin 2\theta_d$ ranges from 0.99 to 0.3 while $\varepsilon_K$ 
takes values between $(1-3) \times 10^{-3}$.
Regarding values of the relevant mass insertions, the values of 
$a,\;b,\;c$ corresponding to the crosses in Fig.~\ref{eps-beta}, 
we find that  
$\sqrt{|\operatorname{Im}\left[(\delta^u_{LL})^2_{13}\right]|}$, 
$\sqrt{|\operatorname{Im}\left[(\delta^u_{RL})^2_{13}\right]|}$ and 
$\sqrt{|\operatorname{Im}\left[(\delta^u_{LL})_{13}\right]
\operatorname{Im}\left[(\delta^u_{RL})_{13}\right]|}$,
are of the same order of maginitude and range from 0.08 to 0.2 
as $c$ varies from -1.5 to 0.5. The mass insertions  
$\sqrt{|\operatorname{Im}\left[(\delta^u_{LL})_{13}\right]
\operatorname{Im}\left[(\delta^u_{LL})_{23}\right]|}$,
$\sqrt{|\operatorname{Im}\left[(\delta^u_{RL})_{13}\right]
\operatorname{Im}\left[(\delta^u_{RL})_{23}\right]|}$,
$\sqrt{|\operatorname{Im}\left[(\delta^u_{LL})_{13}\right]
\operatorname{Im}\left[(\delta^u_{RL})_{23}\right]|}$,
take values from 0.065 to 0.12 in the same range.

We notice that the values of these mass insertions are compatible  
with those given in Ref.~\cite{Gabrielli:2002fr} 
to saturate the SUSY contributions to $a_{J/\psi K_S}$ 
\footnote{
The set of input parameters corresponding to Fig.~\ref{eps-beta} 
are $\av{m_{\tilde{q}}}=534$~GeV, $\mu=478$~GeV, $m_2=210$~GeV 
and the mass of the right-handed stop $m_{\tilde{t}_R}=289$~GeV.}.
As a specific example,  
for $a=b=-3$ and $c=-0.9$ we find 
$\varepsilon_K= 2\times 10^{-3}$, and $\sin 2 \theta_d=0.7$. The most relevant 
mass insertion for the gluino contribution is:
$(\delta^d_{31})_{LL}=-0.0012 -0.066 i$,
being  the other contributions below $10^{-5}$. The relevant 
mass insertions for the chargino contribution are
$(\delta^u_{31})_{LL}=0.0014-0.087 i$, 
$(\delta^u_{31})_{RL}=0.00051-0.11 i$,
$(\delta^u_{32})_{LL}=0.011+0.047 i$ and 
$(\delta^u_{32})_{RL}=-0.018+0.059 i$ . 

Note that although we assume that $A^u=A^d$ at the GUT scale, due to the effect
of the large top Yukawa in the evolution to the electroweak scale, one 
obtains $\operatorname{Im}[(\delta^u_{31})_{RL}]\approx 0.1$ while 
$\operatorname{Im}[(\delta^d_{31})_{RL}]\approx 10^{-5}$. Therefore, the
chargino exchange gives  the dominant contribution to $\sin 2 \theta_d$.

To conclude this section, we address the $\Delta B=1$ process
$b \to s \gamma$. As aforementioned, USY models have large mixing,
hence one might expect that the mass insertions  
$(\delta^d_{23})_{LL}$ and $(\delta^d_{23})_{LR}$ receive large
contributions, and this could lead to excessive values for the
branching ratio (BR) of the $b \to s \gamma$ decay, above the
experimental limit reported by the CLEO collaboration~\cite{bsg:cleo}:
\begin{equation}\label{b:cleo}
2.0 \times 10^{-4} < BR\;(B \rightarrow X_s \gamma)
<4.5 \times 10^{-4} \;.
\end{equation}
The supersymmetric contributions to the process are given by the
one--loop magnetic dipole and chromomagnetic dipole penguin diagrams,
which are mediated by charged Higgs boson, chargino, gluino and
neutralino exchanges. 

It is known~\cite{b:gabrielli} that with universal
$A$-terms, the gluino contribution is very small and the leading ones
are those involving the charged Higgs and charginos. In this case, and
for our choice of USY phases, we found that the total 
$BR\;(B \rightarrow X_s \gamma)\; \simeq 3.5 \times 10^{-4}$, which is
essentially the SM central value. For the case of non-universal
trilinear terms, we present just an illustrative example. Let us
consider (for the structure given in Eq.~(\ref{Aterms})), $a=-1.7, b=c=-3$,
a point in parameter space associated with a nearly maximal value of 
$\sin (2 \beta +2 \theta_d)$. In this case 
$BR\;(B \rightarrow X_s \gamma)\; \simeq 2.8 \times 10^{-4}$, still in
agreeement with SM results, as discussed in Ref.\cite{b:gabrielli}.

%
\section{\bf \large Discussion and Conclusions}

In this work we have studied the implications of having universal
strength of Yukawa couplings within the MSSM. This class of ans\"atze
for the quark mass matrices is specially interesting since it
provides a relationship between the observed pattern of quark
masses and mixings and the possibility of having small CP violating
phases. We have addressed the problem of CP violation in supersymmetric
USY models, showing that the strength of CP violation due to the standard
KM mechanism is too small to account for the observed CP violation, 
and essential contributions from supersymmetry are required. 

We have shown that the trilinear soft terms play a key r\^ole in
embedding USY into SUSY. In fact, due to the large mixing and
associated phases, 
the constraints from the EDMs on the SUSY parameter space are far
more stringent than in the case of a standard Yukawa parametrization.
Although the effect of these phases is absent in the SM
and in SUSY models with flavour conserving soft breaking terms, once we
require a non--universality in the trilinear terms (which proves to be
essential to saturate the CP observables) there are large
contributions to the electric dipole moments of the neutron and
mercury atom.
We found that in order to satisfy the bound of the mercury EDM the
$A$--terms should be matrix factorizable, and their phases constrained
to be of order $10^{-2}-10^{-1}$.

Within the region of the supersymmetric parameter space where one had
compatibility with the EDM measurements, we have investigated the new
contributions to both K and $B$ system CP observables, and 
we have shown that in the present model it is possible to saturate the
experimental values of $\varepsilon_K$ and
$\varepsilon^\prime/\varepsilon$. Gluino mediated boxes with $LL$ mass
insertions provide the leading contributions to $\varepsilon_K$, while 
$\varepsilon^\prime/\varepsilon$ is dominated by chargino loops,
through $LL$ flavour mixing.

Finally, we
considered the CP asymmetry of the $B$ meson decay, $a_{J/\psi K_S}$.
It turns out that within this model the SM contribution to this asymmetry 
is negligible, while supersymmetry, namely through chargino exchange,
provides the leading contributions, which are in agreement with  
the recent measurements at BaBar and Belle.

In conclusion, we have presented an alternative scenario for CP
violation, where the 
strength of CP violation originated from the SM is naturally small, 
due to the observed pattern of quark masses and mixing angles, 
which constrain all USY phases to be small.
We have shown that in this framework new SUSY contributions are 
essential in order to generate the correct value of 
$\varepsilon_K$ and  $\varepsilon^\prime/\varepsilon$, as well as 
the recently observed large value of $a_{J/\psi K_S}$.

%
\section*{\bf \large Acknowlegdments}

S.K. would like to thank CFIF for its kind hospitality during the
final stage of this work. We are grateful to F.~Joaquim and
K.~Tamvakis for useful discussions.
This work was supported in part by
the Portuguese Ministry of Science through project CERN/P/Fis/40134/2000, 
CERN/P/Fis/43793/2001, and by the E.E. through project HPRN-CT-2000-001499.
M.G. and A.T. acknowledge support from 'Fundac\~ao para a Ci\^encia e
Tecnologia', under grants SFRH/BPD/5711/2001
and PRAXIS XXI BD/11030/97, respectively.  
The work of S.K. was supported by PPARC.

%

\section*{\bf \large Appendix: USY and a local horizontal symmetry}\label{app}
In this appendix we show how USY textures can be motivated by
a horizontal symmetry. For instance, with the horizontal gauge group 
$SU(3)_H \times U(1)_H$, the matter content 
of the MSSM can be assigned as $\{ Q_a , L_a \}\equiv (3,1)~\mathrm{and}~ 
\{ u^c_a ,\; d^c_a,\; e^c_a\} \equiv (\bar{3},-1)\;,$
while the charges of the MSSM Higgs are given by $\{ H_u , H_d \} 
\equiv (1,0)$.  The extra Higgs fields that may be used to break 
$SU(3)_H \times U(1)_H$ are $\phi_a \equiv (3,1)$ and 
$\bar{\phi}^a \equiv (\bar{3} ,-1)$, $a=1,2,3$. 
We introduce an additional $Z_4$ discrete symmetry, under which 
$\{ H_u , H_d \}\rightarrow \{ -H_u , -H_d \}$, 
$\{{\phi_a, \bar{\phi}^a }\}\rightarrow \{{i \phi_a, i \bar{\phi}^a \}}$, 
while all other matter superfields transform trivialy. This symmetry 
prevents the presence of renormalizable terms involving 
quarks and leptons in the superpotential. Therefore,
the  lowest dimensional 
$SU(3)_H \times U(1)_H$ invariant
operators in the superpotential, which are 
responsible for generating the fermion masses, are given by
\begin{equation}\tag{A2}
W_{\mathrm{Yuk}} = h_u Q_a u^c_b H_u \frac{\bar{\phi}^a \phi_b}{M^2} + h_d Q_a
d^c_b H_d \frac{\bar{\phi}^a \phi_b}{M^2}+ h_l L_a e^c_b H_d
\frac{\bar{\phi}^a \phi_b}{M^2},
\end{equation}
where $M$ is a scale much higher than the weak scale. 
The minimization of the scalar potential involving the fields 
($\bar{\phi}^a,\ \phi^b$) can be derived in a similar 
way as in Refs.~\cite{Khalil:2002jq,Babu:1999js}.
Thus $ Y^u_{ab} = h_u \frac{\bar{v}^a v^b}{M^2}$, 
where $\bar{v}^a$, $v^b$ are the
vacuum expectation values of $\bar{\phi}^a_i=\bar{v}^a \delta_{ai}$ 
and $\phi^b=v^b\delta_{bi}$ ($i=1,2,3$), respectively. 
In fact, 
for a particular choice of the soft terms associated 
to the fields ($\bar{\phi}^a,\ \phi^b$) one can assume 
that if $v^a$ and $\bar{v}^b$ are given by 
$v^a = \vert v \vert e^{i \varphi_a}$
and $\bar{v}^b = \vert v \vert e^{i \varphi'_b}$, then
\begin{equation}\tag{A3}
Y_{ab}^u = h'_u \;e^{i (\varphi_a + \varphi'_b)}\;, \quad 
\mathrm{with} \quad
h'_u= \frac{h_u {|v|}^2}{M^2}\;.
\end{equation}
Similar expressions hold for $Y^d$ and $Y^l$.
The Yukawa couplings in Eq.~(A3) clearly display the usual USY form, 
which was discussed in section~\ref{usy}.

%

\end{document}